\documentclass[pdflatex,sn-mathphys-num]{sn-jnl}

\usepackage{graphicx}% Include figure files
\usepackage{amsmath,amssymb,amsfonts}%
\usepackage{amsthm}%
\usepackage{svg}
\usepackage{dcolumn}% Align table columns on decimal point
\usepackage{bm}% bold math
\usepackage{siunitx}
\usepackage{hyperref}% add hypertext capabilities

\graphicspath{{./figures/},{../figures}}

\usepackage{soul}
\usepackage{todonotes}

\begin{document}

\title{Demonstration of strong coupling of a subradiant atom array to a cavity vacuum}
%Demonstration of strong collective coupling of atoms to a cavity vacuum via destructive Bragg scattering}% Force line breaks with \\

\author[1,2]{Bence Gábor}
\author[1,3]{K. V. Adwaith}
\author[1,4]{Dániel Varga}
\author[1]{Bálint Sárközi}
\author[1]{Árpád Kurkó}
\author[1]{András Dombi}
\author[1]{Thomas W. Clark}
\author[1]{Francis I. B. Williams}
\author*[1]{David Nagy}
\email{nagy.david@wigner.hun-ren.hu}
\author[1]{András Vukics}
\author[1]{Peter Domokos}

\affil*[1]{HUN-REN Wigner RCP, H-1525 Budapest P.O. Box 49, Hungary}
\affil[2]{Department of Theoretical Physics, University of Szeged,
  Tisza Lajos k\"{o}r\'{u}t 84, H-6720 Szeged, Hungary}
\affil[3]{Université Paris-Saclay, CNRS, ENS Paris-Saclay,
  CentraleSupélec, LuMIn, 91190 Gif-sur-Yvette, France}
\affil[4]{Department of Physics of Complex Systems, ELTE Eötvös Loránd
  University, Pázmány Péter sétány 1/A, H-1117 Budapest, Hungary}

\date{\today}

\abstract{
By considering linear scattering of laser-driven cold atoms inside an undriven high-finesse optical resonator, we experimentally demonstrate effects unique to a strongly coupled vacuum field. Arranging the atoms in an incommensurate lattice with respect to the radiation wavelength, the Bragg scattering into the cavity can be suppressed by destructive interference: the atomic array is subradiant to the cavity mode under transverse illumination. We show however, that strong collective coupling leads to a drastic modification of the excitation spectrum, as evidenced by well-resolved vacuum Rabi splitting in the intensity of the fluctuations. Furthermore, we demonstrate a significant polarization rotation in the linear scattering off the subradiant array via Raman scattering induced by the strongly coupled vacuum field.}

\maketitle

\clearpage

\section{Introduction}

Beyond the peculiar emission from collective
  Dicke-states of an ensemble of indistinguishable atoms
  \cite{gross_superradiance_1982}, the concept of superradiance
  \cite{devoe_observation_1996,inouye_superradiant_1999,araujo_superradiance_2016,ferioli_non-equilibrium_2023}
  and subradiance
  \cite{guerin_subradiance_2016,das_subradiance_2020,rui_subradiant_2020}
  has recently been extended to ordered atom arrays
  \cite{zoubi_excitons_2013,bettles_enhanced_2016} in which the
  interplay of the resonant dipole-dipole interaction together with a
  constructive or destructive spatial interference leads to
  enhancement, or inhibition of spontaneous emission, respectively.
The latter has application in long-term storage of quantum information
\cite{plankensteiner_selective_2015,facchinetti_storing_2016,ferioli_storage_2021}. Accordingly,
there has been an extensive study of regular one-, two-, and
three-dimensional atomic arrays, e.g., subradiance has been shown to
correspond to optical guided polariton modes in the atomic array
\cite{asenjo-garcia_exponential_2017}. Besides ordering atoms, the
radiation can also be shaped to favour emission into selected output
channels, such as when coupling them to fibre-guided modes
\cite{solano_super-radiance_2017,albrecht_subradiant_2019,pennetta_observation_2022}.
Confinement of the electromagnetic field to
  waveguides or resonators also results in a spatial enhancement of
  the range of radiative atom-atom interactions, reinforcing the
  formation of collective states of an atomic ensemble. Subradiant
configurations have been experimentally demonstrated for a
one-dimensional array near a waveguide
\cite{van_loo_photon-mediated_2013}.

In this paper, we revisit low-intensity light scattering from a one-dimensional atom array, when the scattered output is directed into strongly coupled radiation modes sustained by an optical resonator \cite{reimann_cavity-modified_2015,yan_superradiant_2023}.  Dynamics of laser-driven atoms interacting with cavity field modes is of high interest producing a great variety of effects:  experiments started with efficient cooling schemes \cite{nussmann_submicron_2005}, atomic self-organization \cite{domokos_collective_2002,black_observation_2003,arnold_self-organization_2012} and led to the exploration of superradiant \cite{slama_cavity-enhanced_2007,zhang_dicke-model_2018,bohr_collectively_2024} and other types of quantum phase transitions \cite{baumann_dicke_2010,landig_quantum_2016,leonard_supersolid_2017,klinder_observation_2015,kollar_supermode-density-wave-polariton_2017,muniz_exploring_2020,kesler_observation_2021,mivehvar_cavity_2021,PhysRevA.105.063712,PhysRevA.107.023713,helson_density-wave_2023}. Collective radiation effects in many-atom cavity QED systems have been explored, such as the interference in Rayleigh scattering with controlled positioning of atoms in a cavity mode \cite{reimann_cavity-modified_2015,casabone_enhanced_2015,neuzner_interference_2016,yan_superradiant_2023,hotter_cavity_2023}, quantum non-demolition measurements \cite{mekhov_quantum_2009}, as well as lasing \cite{guerin_mechanisms_2008,sawant_lasing_2017} and superradiant lasing \cite{meiser_steady-state_2010,bohnet_steady-state_2012,norcia_superradiance_2016} with cold atoms as the gain media.

We explore the spectral and polarization properties of scattering from a cold atomic ensemble into a quasi-resonant mode of a high-finesse optical cavity. The atoms are arranged into an optical lattice with periodicity incommensurate with the wavelength of the cavity mode resonating with the driven atomic transition. When they are illuminated from a direction perpendicular to the cavity axis, the coherent Bragg scattering from the atom array is suppressed into the cavity.  However, destructive interference does not entail a decoupling from the cavity field even if the laser-driven atom array is without the Bragg condition. There is a collective strong coupling between the subradiant array and the cavity mode, which is manifested by vacuum Rabi splitting \cite{thompson_observation_1992,tuchman_normal-mode_2006,hernandez_vacuum_2007,courteille_photonic_2025}  in the frequency dependence of the outcoupled cavity field intensity fluctuations. We observe another remarkable effect unusual in the fluorescence of atoms in the low-saturation limit: the field polarisation is rotated.  In coherent Rayleigh scattering, the dipole oscillation of an atom is parallel with the polarization of the impinging field; hence, the scattered field preserves this polarization. This component is, however, suppressed by the destructive interference in the atom array subradiant to the cavity field. The incoherent scattering is enhanced in a high-finesse cavity also into the mode with polarization orthogonal to that of the incoming field. The polarization rotation is associated with a two-photon Raman transition in the atomic hyperfine ground state manifold in accordance with the conservation of angular momentum \cite{vrijsen_raman_2011,zhang_dicke-model_2018,suarez_collective_2023}. We show that this process is on the same order of the drive power and reflects the same vacuum Rabi split spectrum as the polarization-preserving scattering.

\section{Results}

\begin{figure}[thbp]
\begin{center}
\includegraphics[width=0.75\columnwidth]{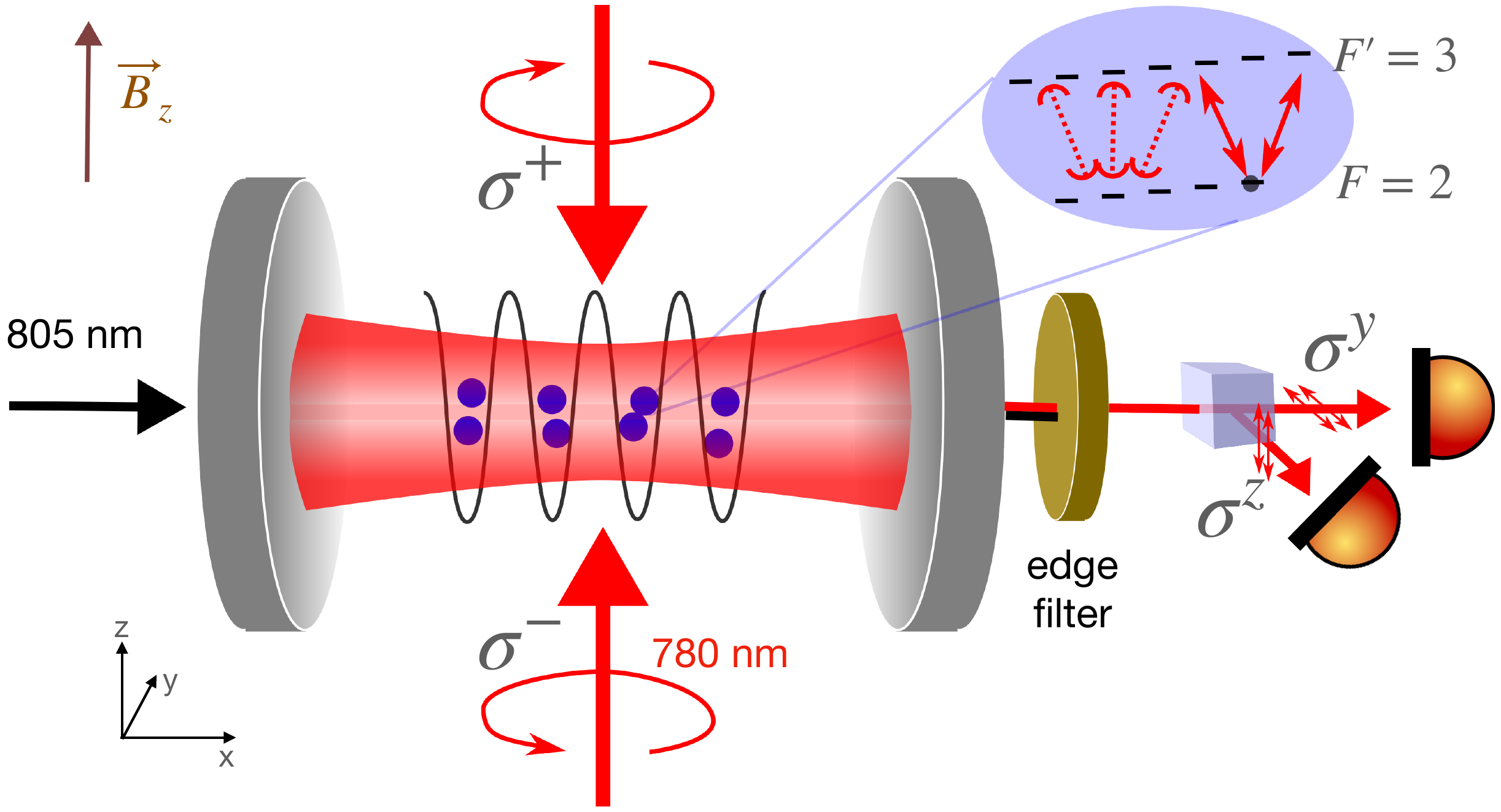}
\caption{Scheme of our experiment on the scattering from a subradiant atomic configuration. Cold ${}^{87}$Rb atoms in an intra-cavity dipole lattice at wavelength 805 nm are illuminated by two counter-propagating coherent laser beams with equal intensity and opposite circular polarizations from the two opposite directions perpendicular to the cavity axis. The laser was set near resonant with the $F=2 \leftrightarrow F'=3$ transition of the D2 line at \SI{780}{nm} (only two transitions from the sublevel $m_F=2$  are shown by solid arrows, for simplicity, but all the other $m_F$ sublevels are coupled similarly by $\sigma^+$ and $\sigma^-$ transitions) and close to resonance with one of the fundamental cavity modes coupling to the atomic transitions denoted by dashed lines in the inset (only three transitions from $m_F=-1$ are shown but all the other sublevels are similarly coupled by the cavity modes). The cavity field output is monitored by single photon counters on discriminating the photon polarization. The cavity linewidth is $\kappa=  2\pi \times \SI{4}{\MHz}$ (HWHM), the maximum single-atom coupling constant is $g= 2\pi \times \SI{0.33}{\MHz}$. }
\label{fig:scheme}
\end{center}
\end{figure}

The experimental scheme is sketched in Fig.~\ref{fig:scheme}. 
Rubidium atoms were trapped in an \SI{805}{nm} optical lattice sustained by a resonantly driven TEM$_{00}$ mode of a high finesse optical cavity \cite{VARGA2024129444}. Another, undriven, fundamental mode of bare cavity resonance frequency $\omega_\text{C}$, was set to resonance with the $F=2 \leftrightarrow F'=3$ atomic excitation frequency $\omega_\text{A}$. The weak probe laser beam, of frequency $\omega$ was swept over about $\pm$\SI{50}{MHz} around $\omega_\text{A}$, and illuminates the atoms from a direction perpendicular to the cavity axis. The common detuning $\Delta = \Delta_\text{A} = \Delta_\text{C}$ where $\Delta_\text{A} \equiv \omega - \omega_\text{A}$ and $\Delta_\text{C}  = \omega - \omega_\text{C}$.  Two single photon counters record separately the cavity output for vertical (z) and horizontal (y) linear polarizations. 

Cavity photons in the mode with frequency $\omega_\text{C} \approx \omega$ could be generated only by scattering from the laser drive beams. Since the atomic distribution had a periodicity incommensurate with the wavelength of the drive (\SI{780}{nm}), the  scattered coherent wave components from different positions of the mode average out along the cavity axis \cite{zippilli_suppression_2004,yan_superradiant_2023}. Quantum emitters can be prepared in sub-radiant states such that the collective emission amplitude is deterministically canceled out, as seen in perfectly ordered atomic arrays \cite{mekhov_cavity_2007} or pure Bose-Einstein condensates. In the present case, the atoms have a finite thermal motion along the cavity axis around the trap centers. Therefore, suppression of the collective coherent scattering into the cavity is expected only on average over a large statistical ensemble. In each individual measurement, density fluctuations in the half wavelength \SI{805}{nm} lattice order lead to cavity field fluctuations that are monitored in the outcoupled field by the photodetectors.

\subsection{Vacuum Rabi splitting}

\begin{figure}
    \centering
    \includegraphics[width=0.7\linewidth]{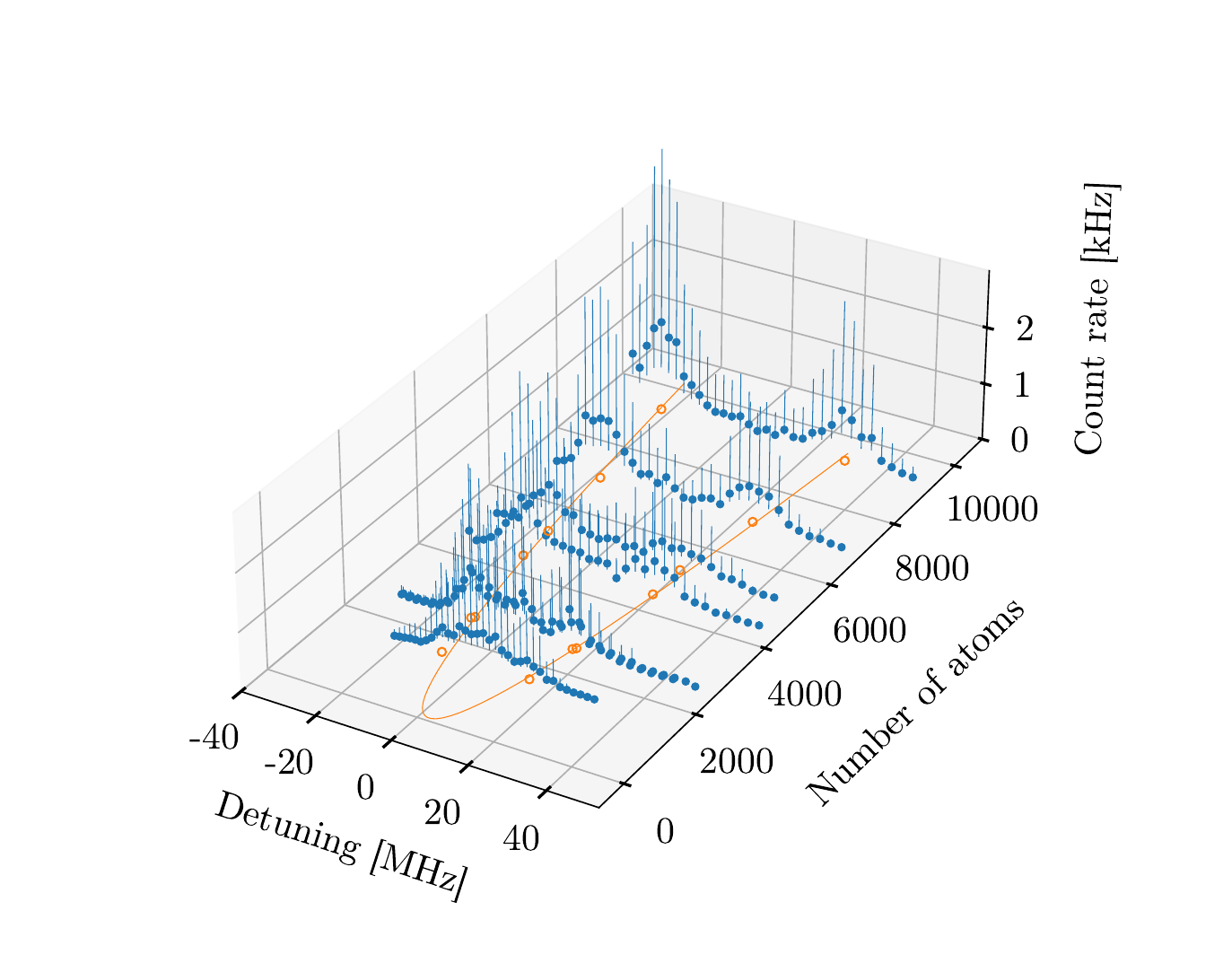}
    \caption{Vacuum Rabi splitting with a subradiant array of
      atoms. The photon count rate in the first \SI{1}{\milli\s} of
      exposure time is plotted versus the laser drive detuning
      $\Delta$ for various effective atom numbers $N_{\rm eff}$. Each
      point and error bar is obtained from an ensemble of 70 runs
      assuming log-normal distribution, given that the photon count
      rate is a priori a non-negative quantity. The maxima of the
      fitted Lorentzian resonance functions,
      projected on the bottom plane (orange circles), fit well on a
      parabola $N_{\rm eff} = \Delta^2/g_\text{eff}^2$ with
      $g_\text{eff}=2\pi\times\SI{0.26}{\MHz}$, in accordance with the
      $\sqrt{N_{\rm eff}}$ dependence known for the collective coupling of a
      number of $N_{\rm eff}$ atoms to a single cavity mode.}
    \label{fig:VRS}
\end{figure}

To start, the number of atoms loaded into the mode volume was varied
in the range of $\sim1500$ to $\sim10^4$ by setting different MOT
cycle protocols. The effective atom number $N_{\rm eff}$ was determined from independent measurements: it was calibrated by the cavity transmission of a near resonant weak probe detuned from the atomic transition such that the atoms acted as a dispersive medium.  The transverse drive laser intensity was lowered as much as possible so that a reasonable rate of photons, $\sim 1000$ count/second, scattered by the atoms into the cavity could be detected by the single photon counters well above the background. It was $\sim 50$ count/second coming from ambient light and the 805 laser, while the intrinsic dark count rate was below 1 count/second for the superconducting
nanowire single-photon detector. So different drive power was employed for different atom numbers to get this required level of photo-detection rate. Then, at fixed atom number and corresponding drive intensity, the drive laser detuning $\Delta$ was varied in the range of $\pm \SI{50}{\MHz}$ to probe the excitation spectrum of the system.  Figure~\ref{fig:VRS} presents that the intensity fluctuations reflected the vacuum Rabi splitting for large enough atom number.
The observed large variance is intrinsic to the density fluctuations
of atoms in a subradiant configuration set by the \SI{805}{nm}
wavelength intra-cavity optical lattice. As there is no perfect
destructive interference for the finite-size sample of atoms, there is
a field with random amplitude. The shot-to-shot fluctuations of the
field intensity was found close to the average, indicating chaotic
light statistics. The main features of the spectra fit a sum of four
Lorentzians (see below at Fig.~\ref{fig:VRS_pump}), 
  the outer resonances corresponding to normal mode splitting, the
  inner ones are relevant only for the smallest atom number. The
  positions of these outer two Lorentzian peaks projected onto the
  detuning-atom number plane in Fig.~\ref{fig:VRS} are well described
by a parabola reflecting the $\Delta \propto \sqrt{N_{\rm eff}}$
expected for strong collective coupling. The coefficient $g_{\rm eff}
\approx 2\pi\times\SI{0.26}{\MHz}$ from the fit {(with uncertainty
  $2\pi\times\SI{0.006}{\MHz}$)} is in good agreement with the
expected value of $2\pi\times\SI{0.225}{\MHz}$ which can be obtained
by averaging over the atomic population distributed evenly in the
$F=2, m_F$ magnetic sublevels with the corresponding Clebsch-Gordan
coefficients. We attribute the 10\% deviation to the small but not
entirely negligible saturation in the atom number calibration
measurement.

\subsection{Linear scattering}

\begin{figure*}
    \centering
    \includegraphics[width=.48\linewidth]{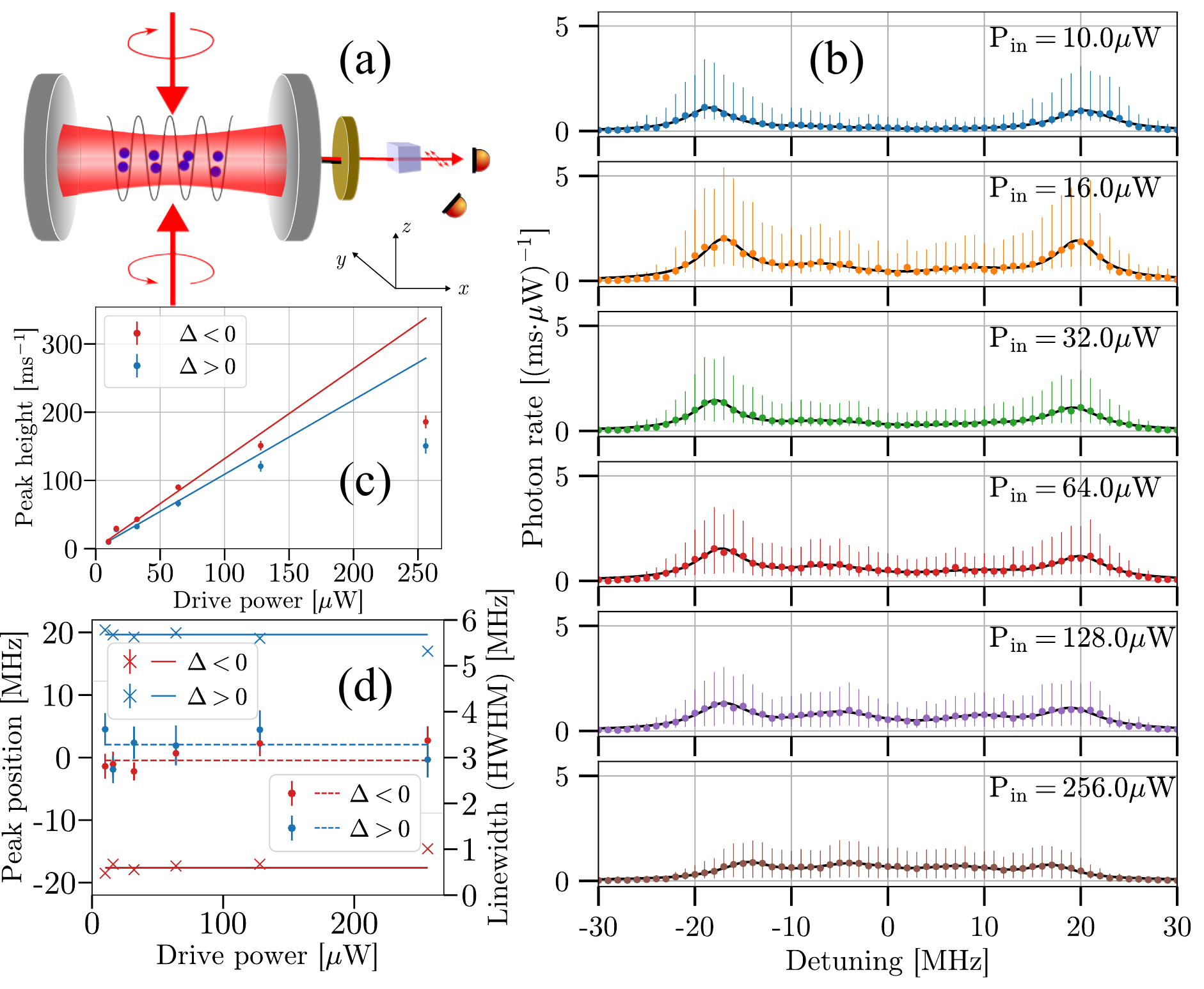} \hfill
    \includegraphics[width=.48\linewidth]{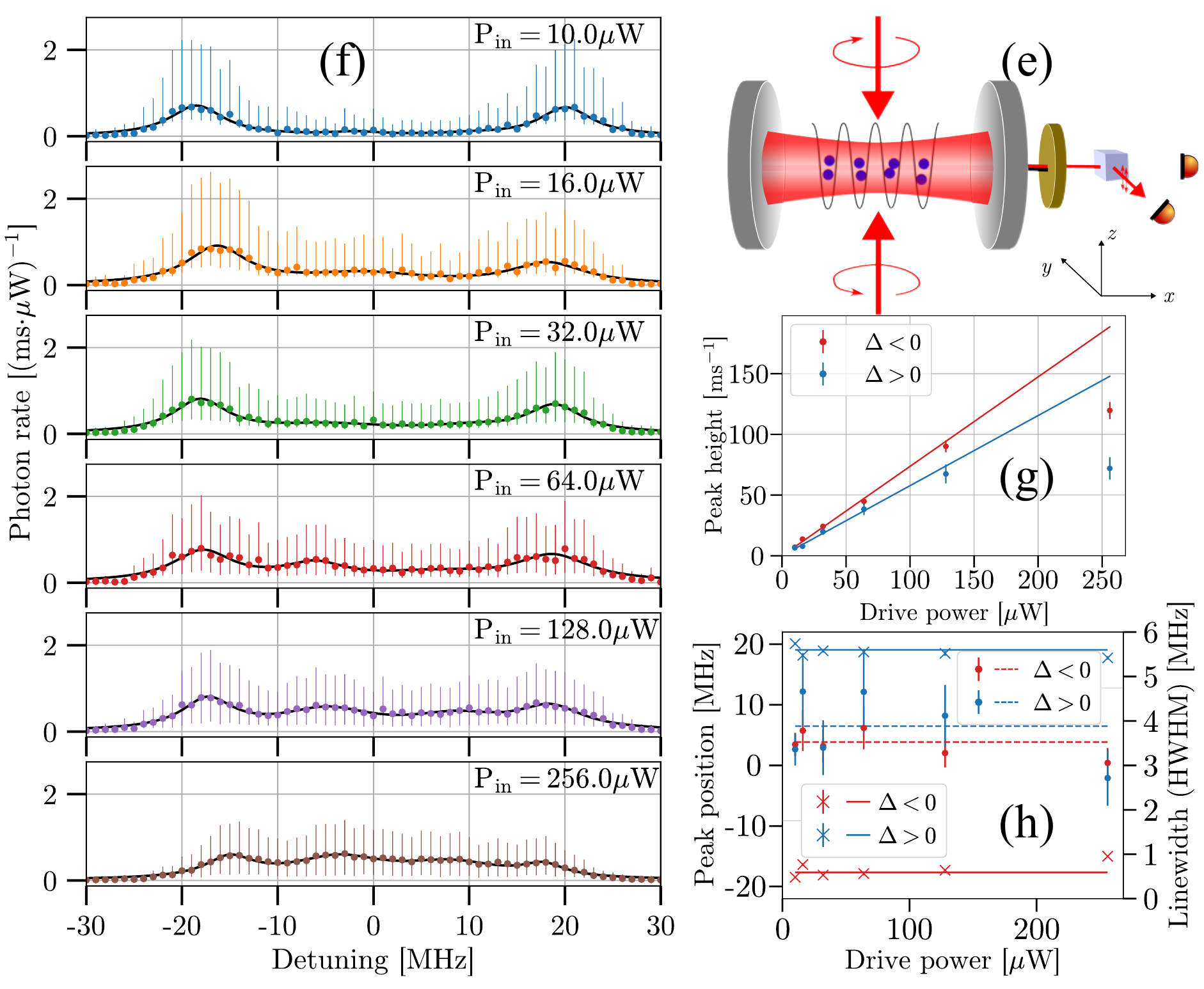}
    \caption{{Power dependence of the vacuum Rabi splitting spectrum for horizontal (a-d) and vertical (e-h) polarization. In panels (b) and (f),  the photon count rate normalized to the laser drive power as a function of the drive frequency is shown for 10, 16, 32, 64, 128 and \SI{256}{\micro\watt} in the subplots from top to bottom. Each point is obtained by averaging 50 runs of  \SI{100}{\micro\s} exposure time, and the statistical variance is represented by the error bars. A sum of four Lorentzian curves is a very good fit on all the measured spectra (see text),  shown by solid lines. The photon count rate at the outer two peaks of the fit spectra are shown in panels (c) and (g), whereas the corresponding detunings are shown in panels (d) and (h) (left scale, red and blue crosses for negative and positive detunings, respectively). The linewidths of the vacuum Rabi peaks are also presented in panels (d) and (h) (right scale, dots). There is a constant fit on the linewidth data (dashed lines, red and blue, according to the sign of detuning) and similarly on the peak positions (solid line) evidencing the linear scattering regime for the drive power range below \SI{128}{\micro\watt}. At \SI{256}{\micro\watt}  saturation effects can be noticed: the peak heights deviate from the linear dependence, and the splitting between the Rabi peaks is also smaller (Data points from the \SI{256}{\micro\watt} measurement are not included in the fits in (c), (d), (g), (h)).}}
%    \caption{{Power dependence of the vacuum Rabi splitting spectrum for horizontal (a-d). In panel (b), the photon count rate normalized to the laser drive power as a function of the drive frequency is shown for 10, 16, 32, 64, 128 and \SI{256}{\micro\watt} in the subplots from top to bottom. Each point is obtained by averaging 50 runs of  \SI{100}{\micro\s} exposure time, and the statistical variance is represented by the error bars. A sum of four Lorentzian curves is a very good fit on all the measured spectra (see text),  \textcolor{blue}{shown by solid lines}. The photon count rate at the outer two peaks of the fit spectra are shown in panels (c), whereas the corresponding detunings are shown in panels (d) (left scale, red and blue crosses for negative and positive detunings, respectively). The linewidths of the vacuum Rabi peaks are also presented in panel (d)(right scale, dots). There is a constant fit on the linewidth data (dashed lines, red and blue, according to the sign of detuning) and similarly on the peak positions (solid line) evidencing the linear scattering regime for the drive power range below \SI{128}{\micro\watt}. At \SI{256}{\micro\watt}  saturation effects can be noticed: the peak heights deviate from the linear dependence, and the splitting between the Rabi peaks is also smaller (Data points from the \SI{256}{\micro\watt} measurement are not included in the fits in (c), (d)).}}
    \label{fig:VRS_pump}
\end{figure*}

Analysis of the drive power dependence of the vacuum Rabi splitting confirms that the scattering is in the linear regime.  The recorded spectra could be compared to a simple theory based on a linear polarizability model of atoms \cite{tanji-suzuki_interaction_2011} which assumes that the atomic induced dipole is proportional to the local electric field, $\vec{d} \propto \epsilon_0 \chi(\omega) \vec{E}(\vec{r})$ in the low-excitation limit \cite{ritsch_cold_2013}. 

In our drive field configuration, the two counterpropagating  beams have opposite circular polarizations. The resulting electric field is linearly polarized in a helical pattern along the drive axis `z', i.e., $ \vec{E}(\vec{r})  \| \vec{e}_y \cos{k z} + \vec{e}_x \sin{k z}$. Note that the optical resonator does not sustain modes with $\vec{e}_x$ polarization, being the direction of the cavity axis; hence, effectively, only the linear polarization $\vec{e}_y$ couples into the resonator field.  Linear scatterers lead then to the intracavity field amplitude for the mode polarized in the direction `y' \cite{ritsch_cold_2013,yan_superradiant_2023}
\begin{equation}
\label{eq:meanfield_y}
\alpha_y= \frac{\eta \, g \sum_a \cos{k x_a} \cos{k z_a}}{(i\Delta_\text{A} - \gamma)(i\Delta_\text{C}- \kappa) + g^2  \sum_a \cos^2{k x_a}} \,,
\end{equation} 
where  $\eta$ is an effective drive amplitude and the summation goes over the atoms indexed by $a=1\ldots N$ with positions $\vec{r}_a=(x_a, y_a, z_a)$. The squared modulus of the denominator has two minima which, for our setting of resonance between the atoms and the mode, $\Delta_\text{A}=\Delta_\text{C} = \Delta$, are at $\Delta = \pm \sqrt{g^2  \sum_a \cos^2{k x_a}} \equiv \pm \sqrt{N_{\rm eff}} \, g$. The effective atom number is around $N_{\rm eff} \approx N/2$ for $\overline{\cos^2{k x}} = 1/2$. This two-peaked resonance behaviour is responsible for the normal mode splitting shown in Fig.~\ref{fig:VRS}.  A destructive interference leads to vanishing mean field, $\overline{\alpha}$, which is formally represented by the numerator averaging out over the atomic positions,  $\left\langle \sum_a \cos{k x_a} \cos{k z_a} \right\rangle = 0$. This is the case for a homogeneous distribution, but also for a set of positions $\{x_a\}$ sampling the  \SI{805}{nm} wavelength optical lattice.  Even if the mean vanishes, however, there are finite size and thermal fluctuations of the atomic distribution which result in cavity field intensity fluctuations,  $\overline{|\Delta\alpha_y|^2} \neq 0$. Considering the atomic positions as random variables, the statistical average gives  
\begin{equation}
\label{eq:noise}
\left\langle \left| \sum_{a=1}^N \cos{k x_a} \cos{k z_a} \right|^2 \right\rangle \approx N^\beta/4\; ,
\end{equation}
where the power law scaling with the atom number $N$ encapsulates two generic cases, i.e., the uniform random or perfectly ordered distributions,  leading to $\beta=1$ linear or $\beta=2$ quadratic dependences, respectively. The actual value of the exponent $\beta$ can be deduced from our measured data and gives information on the atomic distribution. For destructive interference, as in our situation where the distribution of atoms is incommensurate with the \SI{780}{nm} $\cos{k x}$ mode function,  $\beta=1$ is expected.  $\beta = 2$  would indicate superradiance with perfect constructive interference. The photon count rate is proportional to the intracavity photon number, i.e., the squared modulus of the amplitude in Eq.~(\ref{eq:meanfield_y}), having a statistical average that can be obtained by using Eq.~(\ref{eq:noise}).  In the large vacuum Rabi splitting regime and in leading order of $(\kappa^2 + \gamma^2)/ N_{\rm eff} g^2 \ll 1 $, {the mean of the intensity fluctuations} can be approximated around the peak maxima by the Lorentzian functions
\begin{equation}
\label{eq:VRS_Lorentz}
\langle |\alpha_y|^2 \rangle \approx \frac{\eta^2\, N^{\beta-1}}{8} \, \left[(\Delta \pm \sqrt{N_{\rm eff}}\, g)^2 + \left(\frac{\kappa+\gamma}{2}\right)^2 \right]^{-1}\,.
\end{equation} 
This form of the Rabi splitting peaks can be tested experimentally to verify the linear polarizability model of atoms. Moreover, this is a crucial result because it provides a direct measure of $\beta$ via the scaling of the peak intensity with the number of atoms $N$.

Figure~\ref{fig:VRS_pump} shows the detected photo-count rate normalized to the pump power.  Solid lines show that a function composed of the sum of four Lorentzian functions is a very good fit to the measured points. The outer two peaks correspond to the vacuum Rabi resonances given by Eq.~(\ref{eq:VRS_Lorentz}). The inner two (smaller) peaks are due to the multiplett structure of the hyperfine states and are significant in the fit to account for the non-vanishing photo-count rate around zero detuning. These peaks  will be studied systematically in a subsequent paper. Here we focus on  the outer two peaks, i.e., the measured vacuum Rabi peaks which have three features substantiating the validity of the model Eq.~(\ref{eq:VRS_Lorentz}). First, their separation is constant in the range of pump powers investigated, c.f.~Fig.~\ref{fig:VRS_pump}(d).  It follows then that no noticeable atomic saturation takes place apart from the strongest drive plotted. As a by-product,  this peak separation can be used to calibrate $N_{\rm eff}$. Second, the peak heights of the curves are proportional to the drive power, $\eta^2$, which is shown in Fig.~\ref{fig:VRS_pump}(c).  Some tendency of shrinking peak separation and decreasing peak height can be observed for the strongest drive plotted (\SI{256}{\micro\watt}), which indicates that atomic saturation becomes noticeable at this power. However,  in the power range up to \SI{100}{\micro\watt}, the scattering is clearly in the linear regime. Third, the linewidths of the vacuum Rabi peaks are constant and are close to the theoretical value $(\kappa+\gamma)/2 \approx 2\pi \times 3.5$ MHz. 

\subsection{Subradiant atomic array}

\begin{figure}
    \centering
    \includegraphics[width=0.55\linewidth]{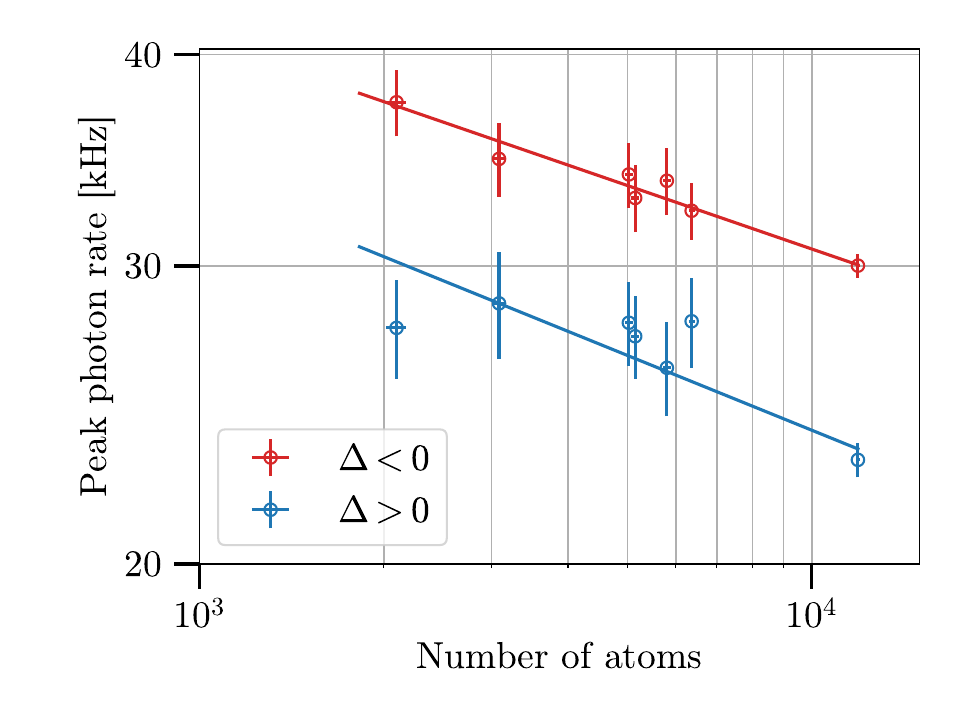}
    \caption{Collective scattering as a function of the atom number. The maximum photon scattering rates on resonance with the vacuum Rabi peaks, both at the negative (red) and positive (blue) side of the detuning, have been detected during a time duration of 100 $\mu$s at a controlled delay after loading the atoms into the lattice. For each atom number, the drive frequency were set to resonance with the vacuum Rabi peaks. Horizontal and vertical polarizations are summed up. The exponents of the linear fit on the log-log scale are obtained $\beta= 0.875 \pm 0.009$ for the red and $\beta =0.853 \pm 0.031$ for the blue, respectively.}
    \label{fig:intensity_vs_atomnumber}
\end{figure}

Having established the linearity of the scattering with driving power, we investigated the dependence of the photon fluctuations scattered into the cavity as a function of the atom number. It was changed by systematically delaying the switch-on time of the transverse drive laser. The drive power was set to a low value, \SI{16}{\micro\watt}, being in the linear scattering regime. The maximum cavity photon number is estimated to be $0.014$ corresponding to a saturation around 1.5\%.  The registered photo-counts were integrated over only  $\SI{100}{\micro\s}$,  in order to minimize the effects of atom loss and atomic motion.  The drive frequency was tuned over the same range as in Fig.~\ref{fig:VRS_pump} so that the full excitation spectrum was recorded. This allowed us (i) to calibrate the atom number from the distance of the peak maxima, and (ii) to determine the peak photo-count rate for the given atom number. At the detunings corresponding to the Rabi resonances, $\Delta = \pm \sqrt{N_{\rm eff}}\, g$, the photo-count rate was measured by averaging over 100 repetitions and is plotted in Fig.~\ref{fig:intensity_vs_atomnumber}. This peak scattering rate can be compared with the theoretical maximum rate at vanishing detuning in the denominator of Eq.~(\ref{eq:VRS_Lorentz}). The corresponding photo-count rate is
\begin{equation}
\frac{I_\text{detect}}{P_\mathrm{dr}} = \xi \, \kappa_T \times
\frac{1}{8}\, \frac{3 \lambda^2}{2\pi {\cal A}_\mathrm{dr} } \times
\frac{\gamma}{\left(\frac{\kappa + \gamma}{2}\right)^2} \times
\frac{1}{\hbar\omega} \approx 6000 \,
\frac{\mathrm{counts}}{\mathrm{s} \cdot \mu\mathrm{W}}\,,
\end{equation}
where $\kappa_T=2 \pi \times 1.15$ MHz is the cavity field decay by mirror transmission, $\xi = 0.5$ is the detection efficiency, $\gamma$ is the atomic linewidth (HWHM), ${\cal A}_\mathrm{dr}$ is the drive beam cross section, and the drive power $P_\mathrm{dr}$ is measured in $\mu$W. With our experimental parameters, the expected count rate is 96 kHz, which is in acceptable agreement with the measured values in the range $30 \pm 3$ kHz, shown in Fig.~\ref{fig:intensity_vs_atomnumber}, provided we take into account some uncontrolled misalignment of the transverse drive laser and incomplete illumination of the entire atom cloud in the cavity. The measured maximum rates scatter within 10\% around a constant value, and show some dependence as a function of atom number, but the power law fit results in an exponent only slightly below 1, which is a consistent with $\beta=1$ in Eq.~(\ref{eq:VRS_Lorentz}). 
This confirms the absence of coherent component in the scattered photon field and supports the observation of subradiance from an array of atoms. Beyond a simplified one-dimensional form of subradiance, the cavity does not merely enhance the scattering into a small solid angle for each individual atom, but the collective strong coupling to the cavity mode modifies the excitation spectrum of the atom array. 

\subsection{Polarization rotation}

%\begin{figure*}
%    \centering
%    \includegraphics[width=.85\linewidth]{Fig3e-h}
%    \caption{Power dependence of the vacuum Rabi splitting spectrum for vertical (e-h) polarization for different transverse laser powers. Parameters are the same as in Fig.~\ref{fig:VRS_pump}. Each point is obtained by averaging 50 runs of  \SI{100}{\micro\s} exposure time, and the statistical variance is represented by the error bars. A sum of four Lorentzian curves is a very good fit on all the measured spectra (see text),  \textcolor{blue}{shown by solid lines}. The photon count rate at the outer two peaks of the fit spectra are shown in panel (g), whereas the corresponding detunings are shown in panel (h) (left scale, red and blue crosses for negative and positive detunings, respectively). The linewidths of the vacuum Rabi peaks are also presented in panel (h) (right scale, dots).}
%    \label{fig:VRS_pump_b}
%\end{figure*}

The strongly coupled vacuum field influences not only the spectral features of scattering but also the polarization. Within the scalar linear polarizability model leading to Eq.~(\ref{eq:meanfield_y}), the atomic polarization induced by the `z' travelling $\sigma^\pm$ beams excites the `y' polarized mode of the cavity. This is Rayleigh scattering which corresponds to the resonance fluorescence of two-level atoms in the low-excitation limit. As far as coherent scattering and interference of the scattered light off many atoms  is concerned, the linear polarizability approach is suitable. However, the degeneracy of the atomic $F=2$ ground-state level changes substantially the incoherent scattering.  The drive excites $(2,m_F) \rightarrow (3,m_F\pm1)$ transitions which have dipole moments in the `y' and `x' directions. The Rayleigh scattering involves thus $(2,m_F) \leftrightarrow (3,m_F\pm1)$ atomic transitions within a two-level system, while creating an `y' polarized photon from the laser drive. The strongly coupled `z' polarized mode of the cavity, however, even in vacuum state, can stimulate $(3,m_F\pm1) \rightarrow (2,m_F\pm1)$ transitions. This is Raman scattering: the initial and final atomic states $(2,m_F)$ and $(2,m_F\pm1)$, respectively, are different along the excitation path. The change of the angular momentum state of the atom compensates for the rotation of the field polarization when a `z' polarized photon is created in the cavity from a field polarized in the $(x,y)$ plane.  Another consequence of the change of the atomic state in the Raman scattering process is that the scattered photon carries which-way information; hence the scattered components from different atoms in the ensemble do not interfere. Regardless the position of the atoms, there is no collective enhancement, nor destructive interference, only the intensities from individual sources add up.

The cavity-induced Raman scattering process has been observed by a photon flux emanating from the mode with polarization `z' ($\sigma_0$) which is the direction of propagation of the input field. The scattering rate into the polarization `z' as a function of the laser drive detuning for a range of drive powers is shown in Fig.~\ref{fig:VRS_pump}(f). It shows very similar features and count rates to the `y' polarization output: (i) vacuum Rabi peaks are linear in the input power (panel (g)) and (ii) constant widths have been measured (panel (h)).  The results show that the cavity-stimulated Raman scattering is also linear in the drive intensity in the low-excitation limit. {This is at variance with the case of two-level atoms where the incoherent part of the scattered light is connected to saturation and is of quadratic order in the drive intensity.} Furthermore, the peak heights were obtained close to those of the polarization-maintaining light scattering, which reinforces the observation that the coherent scattering was strongly suppressed, i.e., another indirect evidence for the subradiance.

The two-photon Raman transition has been exploited to realize quantum interfaces between light polarization and atomic memory states  \cite{wilk_single-atom_2007,specht_single-atom_2011} by means of stimulated adiabatic passage processes with pulsed excitation in single-atom strong-coupling cavity QED experiments. Cavity-enhanced Raman scattering has also been observed from a regular half-wavelength ordered array \cite{yan_superradiant_2023} when the drive is detuned from the atoms. In our experiment, we revealed that the Raman scattering, though being an incoherent process, manifests the vacuum Rabi split spectrum characteristic of the strong collective coupling of the atoms to the `z' polarized cavity mode. 

\section{Discussion}

An important conclusion is that radiation from atomic arrays is not only efficiently collected but is substantially \emph{modified} by the presence of a high-finesse resonator. Most importantly, the strong coupling to selected resonator modes  imposes a \emph{collective} scattering from the atoms into the resonator. This collective coupling, as we have shown, is due to more than simply an interference effect, even in the extremely low intensity limit. The role of collective coupling has been revealed in detecting the collective phase shift of Bragg back-scattered light from a one-dimensional optical lattice along the axis of a ring cavity \cite{slama_phase-sensitive_2005}. Here we realized an experiment where the input field impinges on a one-dimensional atom array from a direction perpendicular to the axis of an undriven resonator, and the collective effect is captured by the Rabi splitting in the intensity of fluctuations around a zero mean-field.

A natural continuation of this work consists in the exploration of the non-linear regime arising at increased powers, e.g., the systematic study of the inner two Lorentzian resonances appearing in the measured data. On a longer time scale, our experiment can be developed toward the realization of new variants of the Dicke model \cite{zhiqiang_nonequilibrium_2017} in disordered manifolds with cavity-mediated interactions. Further,  considering that multiply excited subradiant states are said to be composed of the superposition of singly excited states in random ensembles \cite{ferioli_storage_2021}, our system could be used to provide further insight into this superposition.  In particular, our system is well suited for time-resolved measurements and so the dynamics of the underlying subradiant states in the single-mode limit are available. 

We must conclude, too, that optical polarization enters the linear scattering regime. On the one hand, the multiple ground-state level structure of atoms has to be taken into account in a linear polarizability description of  atoms, beyond the usual scalar polarizability, which was noted as a subtlety in Ref.~\cite{asenjo-garcia_exponential_2017}. On the other hand, the multiple ground states open the possibility of entanglement-based, new type of subradiant states predicted recently \cite{hebenstreit_subradiance_2017}.  In our future work, the cavity-enhanced polarization rotation could be the design basis for  long-range many-body interactions between atoms mediated by two-mode fields. The cavity field fluctuations reflecting a non-trivial atom-cavity spectrum can be exploited as a useful light source when the mean-field is suppressed. Finally, our configuration is very close to schemes for superradiant lasing \cite{bohnet_steady-state_2012} and for atomic clocks \cite{bohr_collectively_2024} which we hope to explore with the incommensurate lattice trap.

\section{Methods}

\paragraph*{Loading atoms into the cavity}

An ensemble of cold ${}^{87}$Rb atoms was collected in a magneto-optical trap (MOT). After the MOT cycle, the atoms were cooled by polarization gradient cooling ($\sigma^+-\sigma^-$ configuration) down to temperatures of \qtyrange[range-units=single,range-phrase=--]{20}{50}{\micro\K}. Subsequently, they are magnetically polarized by optical pumping into the $(F, m_F)=(2, 2)$ hyperfine ground state to allow capture with a magnetic quadrupole trap. The magnetically trapped atomic cloud was then transported into the mode of a high-finesse ($\mathcal{F}/\pi=1430$) resonator by adiabatically displacing the trap center, and was released there by turning off the magnetic field in \SI{7}{\milli\s}.

The cavity is $l=\SI{15}{mm}$ long and the mode waist is $w=\SI{127}{\micro\m}$. A far red detuned (\SI{805}{\nm}) laser beam was injected into the cavity, serving two purposes: firstly, utilizing the Pound-Drever-Hall technique, the cavity was locked to it, secondly, it provided a far-red-detuned optical dipole lattice for the atoms with a depth of \SI{140}{\mu\K} and an unperturbed lifetime of \SI{200}{\milli\s}.

Upon the arrival and release of the atoms, a second optical pumping was performed into the $(F, m_F)=(2, 2)$ hyperfine ground state defined by a homogeneous magnetic field of 1G along `z'. Starting from the initial Zeeman sublevel, the  population spreads over all other sublevels and tends to some steady state population distribution as a result of the competition between fluorescence, Raman transitions and repumping from the $F=1$ level.  The actual steady-state distribution in the $m_F$ sublevels is unknown, but it is not very far from a uniform one (Markov chain simulation according to the Clebsch-Gordan coefficients).

For the measurements leading to Fig.~\ref{fig:VRS}, the atom number was varied by applying a delay (in the range \qtyrange[range-units=single,range-phrase=--]{8}{12}{\ms}, which is safely after the decay of magnetic field transients) before the illumination of the atoms was switched on.

\paragraph*{Calibration of the atom number}

The effective number of atoms which coupled to the
  cavity ($N_{\rm eff}$) was measured from the light shift effect. In
  the dispersive limit, where $\Delta_A \gg \gamma$, the atoms shift
  the cavity mode resonance proportional to their number. The laser
  drive, resonant with the empty cavity mode, was set to $\omega -
  \omega_A = \Delta_A = - 2\pi \times 90$MHz detuning from the atomic
  resonance, hence the corresponding total dispersive frequency shift
  was $N_{\rm eff}g^2/\Delta_A$, that was determined directly from a
  transmission measurement. The effective number $N_{\rm eff}$
  includes the reduced coupling strength $g$ away from the axis
  following a Gaussian transverse mode profile (which effect is the
  same for the transverse drive configuration), and the averaging over
  the mode function $\cos(k x)$ along the cavity axis.
  
\paragraph*{Transverse drive}

%The atoms in the cavity mode were driven by a counterpropagating pair of laser beams in the $\sigma^+-\sigma^-$ configuration, from a direction perpendicular to the cavity axis (transverse pump). <-- This is already described in the main text
This laser was phase-locked to a reference laser with a variable detuning from the atomic resonance, which we scanned in the frequency range $\pm\SI{30}{\MHz}$. The beam waist was \SI{1}{\mm}, the power in each direction was adjusted from \qtyrange{0.15}{256}{\micro\W} by means of an acousto-optic modulator (AOM). Simultaneously, the $F=1\leftrightarrow F'=2$ transition was also driven resonantly by a repumper laser, in order to keep the atoms in the $F=2\leftrightarrow F'=3$ cycle. The beam waist of the repumper was \SI{12}{\mm}, the power in each direction was \SI{4}{\milli\W}. 

\paragraph*{Detection} The 805 nm component was removed by interference filters from the cavity output beam which was then split by a polarizing beam splitter.  Both the horizontal and vertical polarization beams were coupled into a fibre, connected to a superconducting nanowire single-photon detector (for the measurements shown in Fig. \ref{fig:VRS}) or to a single photon counter module (for the measurements shown in Fig. \ref{fig:VRS_pump} and Fig. \ref{fig:intensity_vs_atomnumber}). The overall detection efficiency was 7\% and 50\%, respectively, including the quantum efficiency and the optical coupling into the detector. Although the superconducting nanowire single photon detector has a very high quantum efficiency ($>99\%$), the photon loss was significant during the optical path from the cavity output to the detector including several fibre couplings. Altogether the total efficiency was low (7\%), therefore we decided to install a new detection system with $65\%$ quantum efficiency detectors but with very high optical coupling (finally we reached 50\% overall detection efficiency). We recorded few millisecond long signals with time resolution $\SI{1}{\micro\s}$.

\paragraph*{Shot-to-shot noise in the atom number}
In the experiment the averaging over many realizations may involve a random variation of the atom number. On taking this into account, the Eq.~(\ref{eq:VRS_Lorentz}) is modified and the peak intensity scaling on resonance gets a correction:
\begin{equation}
S_{\rm max} (\Delta_{\rm peak}) = \frac{4 \eta^2 N^{\beta-1}}{(\kappa+\gamma)^2} \left(1 - \frac{4 g^2}{(\kappa+\gamma)^2} \, \frac{\delta N^2}{\overline N} \right)\; ,
\end{equation} 
where $\delta N$ is the variance around the mean $\bar N$. The correction is, however, small for the sub-Poissonian atom number statistics in our MOT. 

\section*{Declarations}

\subsection*{Funding}

This research was supported by the Ministry of Culture and Innovation and the National Research, Development and Innovation Office within the Quantum Information National Laboratory of Hungary (Grant No. 2022-2.1.1-NL-2022-00004), and within the ERANET COFUND QuantERA program  (MOCA, 2019-2.1.7-ERA-NET-2022-00041). A. D. acknowledges support from the János Bolyai research scholarship of the Hungarian Academy of Sciences. B. G. acknowledges the support from the \'UNKP-23-3 New National Excellence Program of the Ministry for Culture and Innovation.

\subsection*{Availability of data and materials}
Not applicable.

\subsection*{Competing interests}
The authors declare no competing interests.

\subsection*{Author contributions}
B.G., K.V.A., and D.N. performed the data acquisition, B.G., B.S., D.N, Á.K. and P.D. contributed to the data analysis, all authors contributed to the overall operations of the experiment, discussed the results, and worked together on the manuscript.

%\bibliography{transverse_rabi_EPJ}

\end{document}